\documentclass[a4paper]{article}
\usepackage{Odyssey2024}
\usepackage{epsfig,amssymb,amsmath}
\ninept
\usepackage{multirow}
\usepackage{listings}
\usepackage{amsmath}
\usepackage{url}
\usepackage{xcolor}
\usepackage{booktabs}
\usepackage{makecell}
\usepackage{adjustbox}
\usepackage{amssymb}
\usepackage{pifont}
\newcommand{\cmark}{\ding{51}}%
\newcommand{\xmark}{\ding{55}}%
\usepackage{float}

\title{PixIT: Joint Training of Speaker Diarization and Speech Separation\\ from Real-world Multi-speaker Recordings}
\name{Joonas Kalda$^1$, Cl\'{e}ment Pag\'{e}s$^2$, Ricard Marxer$^3$, Tanel Alumäe$^1$, Herv\'{e} Bredin$^2$}
\address{$^1$Tallinn University of Technology, Estonia\\ $^2$IRIT, Universit\'{e} de Toulouse, CNRS, Toulouse, France\\ $^3$Université de Toulon, Aix Marseille Univ, CNRS, LIS, Toulon, France}

\begin{document}
\emergencystretch 3em
%
\maketitle
\begin{abstract}
A major drawback of supervised speech separation (SSep) systems is their reliance on synthetic data, leading to poor real-world generalization. Mixture invariant training (MixIT) was proposed as an unsupervised alternative that uses real recordings, yet struggles with over-separation and adapting to long-form audio. We introduce PixIT, a joint approach that combines permutation invariant training (PIT) for speaker diarization (SD) and MixIT for SSep. With a small extra requirement of needing SD labels during training, it solves the problem of over-separation and allows stitching local separated sources leveraging existing work on clustering-based neural SD. We measure the quality of the separated sources via applying automatic speech recognition (ASR) systems to them. PixIT boosts the performance of various ASR systems across two meeting corpora both in terms of the speaker-attributed and utterance-based word error rates while not requiring any fine-tuning.
\end{abstract}

\section{Introduction}
\label{sec:intro}
 Speech separation is the task of estimating individual speaker sources from a mixture. It is an important part of automatic speech technologies for meeting recordings as a significant proportion of the speech can be overlapped. Supervised training approaches, mainly permutation invariant training (PIT), have been shown to perform well on few seconds long fully-overlapped synthetic speech mixtures that fit in the memory for the model \cite{luo_conv-tasnet_2019, luoDualpathRNNEfficient2020}. To extend a PIT-based approach to more realistic data, \cite{chenContinuousSpeechSeparation2020} proposed the task of continuous speech separation (CSS). This involves generating long-form separated sources from a continuous audio stream that contains multiple utterances that partially overlap. The standard method for extending PIT-based separation systems to CSS is by applying them on a sliding window and reordering sources in neighboring chunks based on a similarity metric calculated on the overlapped region. In long-form audio, however, the speaker tracking breaks down if a speaker stops speaking for longer than the overlapping portion of the sliding window. 

 Another problem of PIT-based training that remains in CSS approaches is the reliance on clean single-speaker isolated sources for the synthetic mixtures. The supervised approach does not generalize well to real-world data as clean ground truth
 separated reference signals are not available in recordings due to cross-talk. To combat this, mixture invariant training (MixIT) was introduced in \cite{wisdomUnsupervisedSoundSeparation2020a}, an unsupervised method that does not require clean separated sources for training. Two mixtures from the target domain are added together to obtain a mixture of mixtures (MoM) and a separation model is trained to estimate sources so that they can be combined to obtain the original mixtures. In \cite{sivaramanAdaptingSpeechSeparation2021} it was demonstrated that this method is effective in using real-world meetings as the target domain. A limitation of MixIT is that the number of output sources for the separation model has to be twice the maximum number of speakers of a single mixture. This can lead to over-separation and makes it difficult to generalize to long-form audio. Over-separation can be mitigated by performing semi-supervised training but this still relies on synthetic data. In \cite{liSelfSupervisedLearningBasedSource2023}, MixIT was used in combination with speaker diarization pre-processing to perform source separation on real-world long-form meeting audio. Separation was done at the utterance level and the correct speaker sources to use were determined by comparing speaker embeddings with global embeddings obtained from diarization. This resulted in superior speaker-attributed automatic speech recognition (ASR) performance. A limitation of this approach is the need for extra voice activity detection (VAD) and speaker diarization models to segment long-form audio into speaker-attributed utterances, as speech separation is performed solely at the utterance level.

Traditional speaker diarization approaches have relied on a multi-step approach consisting of VAD to obtain speaker segments, local speaker embeddings, and clustering \cite{landini2022bayesian}. End-to-end diarization (EEND) is a newer approach that is able to handle overlapped speech but comes with its own limitations, such as needing a large amount of data and mispredicting the number of speakers \cite{fujitaEndtoEndNeuralSpeaker2019, fujitaEndtoEndNeuralSpeaker2019a}. Recently the two approaches have been combined into the \textit{best-of-both-worlds} framework \cite{kinoshitaAdvancesIntegrationEndtoend2021a, kinoshitaIntegratingEndtoendNeural2021b} which performs EEND on small chunks and stitches the results together using speaker embeddings and clustering. 

Speech separation and speaker diarization are both often parts of multi-speaker automatic transcription systems. The models used to carry out these two tasks are mostly cascaded in two different ways. Since the sources extracted by a speaker separation system no longer have speech overlap regions, they can greatly facilitate the speaker diarization task improving its performance. An example of such a system is the speaker separation guided diarization system (SSGD) \cite{fang_deep_2021, morrone_low-latency_2023}. A drawback of this method is that diarization depends on the quality of the separated sources. Another option is to place a diarization system upstream of a speaker separation system, like in \cite{boeddeker_ts-sep_2023, raj_gpu-accelerated_2023}. Indeed, source separation is easier if the speech activity of each speaker is known, provided that the diarization system is able to manage speech overlap. Similarly to the previous approach, the speech separation performance depends on the quality of the speaker diarization. Thus, we can see that these two tasks can benefit from the results of the other, highlighting their interdependence, and the fact that there is no obvious choice whether to start the processing with a diarization or speech separation system. This has served as motivation for joint learning approaches. The Recurrent Selective Attention Network architecture (RSAN) \cite{von_neumann_all-neural_2019} was the first all-neural model to jointly perform the speech separation, speaker diarization, and speaker counting tasks. In this model, the extraction is made over time using sliding blocks. In each block, speakers are iteratively extracted from the mixture by estimating a mask for each of them, given speaker embeddings determined in the previous blocks, and a residual mask from the previous iterations in the current block. Another architecture that performs jointly these three tasks is the end-to-end neural diarization and speech separation architecture (EEND-SS) \cite{maitiEENDSSJointEndtoEnd2022}. This system is based on the EEND framework for the diarization and speaker counting tasks and Conv-TasNet \cite{luo_conv-tasnet_2019} for the speaker separation one. In the EEND-SS architecture, the information given by the diarization branch is used to refine the separation part, by providing an estimation of the number of speakers and using the probability of speech activity to enhance the separated source signals. These joint approaches, however, still all rely on synthetic data for separation training. 

We propose a joint framework for performing both speaker diarization and speech separation on long-form real-world audio. We name the approach PixIT, as it combines PIT for speaker diarization and MixIT for speech separation. We leverage speaker diarization information that is often available for meeting corpora to create MoMs that have the maximum number of speakers limited to better mimic real-world mixtures. Our separation/diarization model processes the mixture/MoM and outputs separated source predictions and the respective speaker activity predictions. When training the joint model we combine the PIT-loss for both the original mixtures and MoMs with the MixIT loss for the MoM. Aligning speaker sources with the speaker activations also solves the over-separation problem of MixIT. In inference, we are able to stitch together the separated sources across the sliding windows by first stitching the speaker activations as is done in the \textit{best-of-both-worlds} approach for diarization. To measure the quality of the long-form stitched separated sources, we feed them into a variety of off-the-shelf ASR systems. We observe improvements over the baseline method of speaker attribution done through diarization for all ASR systems and two real-word meeting datasets: AMI~\cite{carletta2005ami} and AliMeeting~\cite{yu_m2met_2022}. Furthermore, we show that when the speaker-attributed transcripts are combined into a single output, the utterance-wise word error rate (uWER) improves.

\section{Joint model}

\begin{figure}[t]
  \centering
  \includegraphics[width=0.7\linewidth]{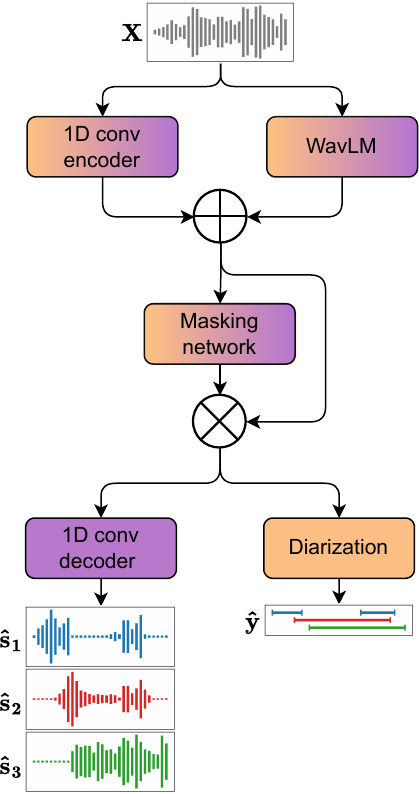}
  \caption{The architecture of the proposed ToTaToNet model.}
  \label{fig:model}
\end{figure}

We base our model on the TasNet architecture \cite{luoTasNetTimedomainAudio2018}, which consists of a 1-D convolutional encoder, a separator module that predicts $N$ masking matrices and a 1-D convolutional decoder. We additionally leverage pre-trained WavLM features \cite{chenWavLMLargeScaleSelfSupervised2022} which are especially suited for speech separation due to the use of the utterance mixing augmentation in their pre-training. These are concatenated with the convolutional encoder outputs. The diarization network takes the encoded separated signals as input and processes each source independently effectively performing VAD. The independent processing of the sources in the diarization module is required to maintain alignment between the separation outputs and the diarization branches. The joint model architecture, which we call ToTaToNet\footnote{Collaboration between labs in \textbf{To}ulouse, \textbf{Ta}llinn, and \textbf{To}ulon}, is illustrated in Figure \ref{fig:model}. The components of the model related to the {\color{orange} \textbf{diarization}} branch are colored {\color{orange} \textbf{orange}}, the components related to {\color{violet} \textbf{separation}} are colored {\color{violet} \textbf{purple}} and the components used by {\color{orange}\textbf{both}} {\color{purple}\textbf{bran}}{\color{violet}\textbf{ches}} are colored a {\color{orange}\textbf{gr}}{\color{purple}\textbf{adi}}{\color{violet}\textbf{ent}} between the two. This color scheme is kept consistent across all the figures in the paper.

\subsection{Training}

\begin{figure*}[t]
\centering
  \includegraphics[width=0.99\linewidth]{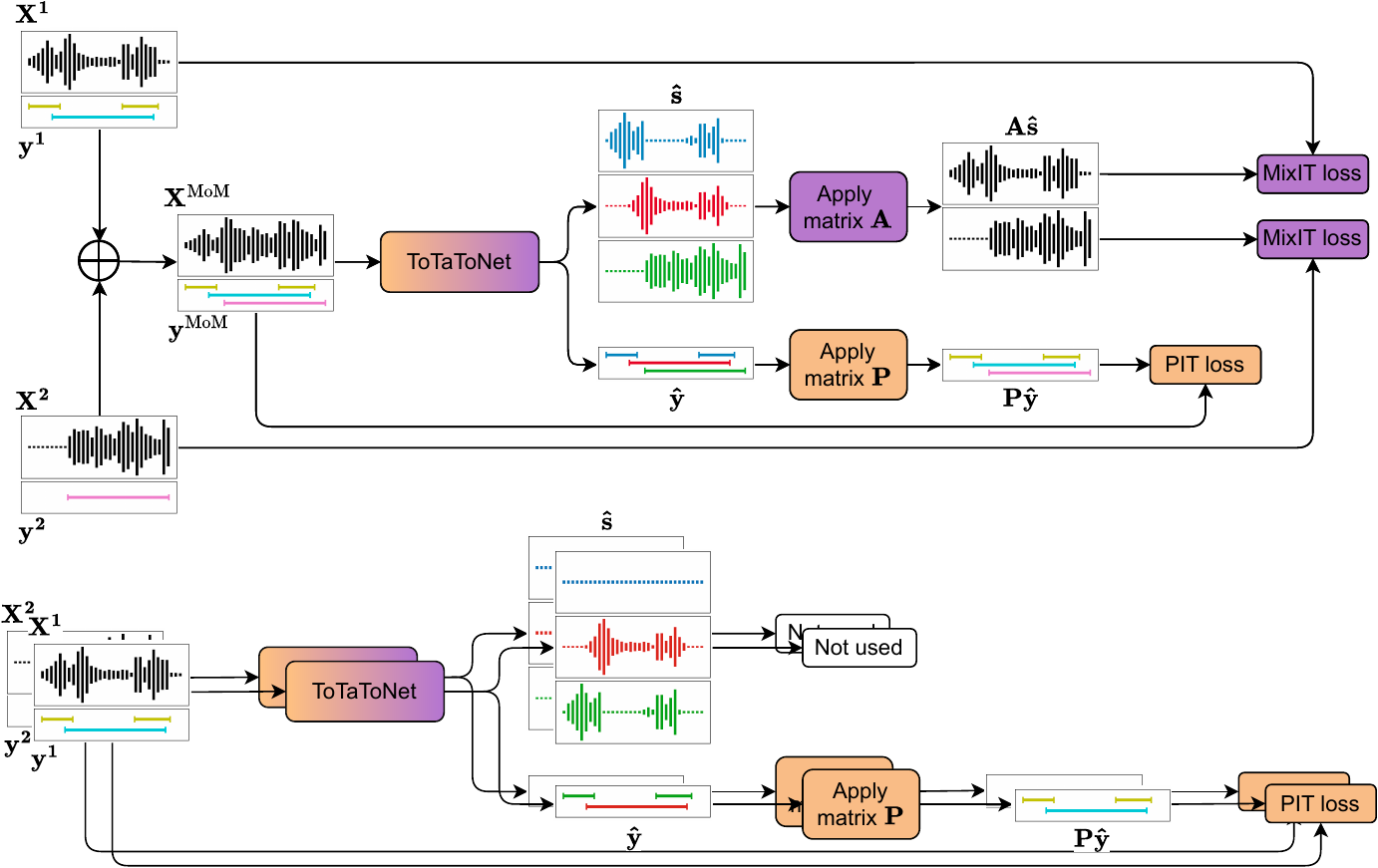}
  \caption{Training the joint model. The upper part shows calculating the MixIT and PIT losses on MoMs. The bottom part shows calculating PIT losses on the original mixtures.}
  \label{fig:training}
\end{figure*}

The joint training method for speech separation and speaker diarization is illustrated in Figure~\ref{fig:training}. Consider an audio chunk $\mathbf{X}$ and the reference speaker activity labels $\mathbf{y} \in\{0,1\}^{K_{\max }\times T}$ where $y_{k,t}=1$ if speaker $k$ is active at frame $t$ and $y_{k,t}=0$ otherwise. Here $K_{\max}$ specifies the maximum number of speakers anticipated in an audio chunk. For diarization, we utilize the well-established permutation-invariant training (PIT) objective \cite{fujitaEndtoEndNeuralSpeaker2019}:
\[
\mathcal{L}_{\mathrm{PIT}}(\mathbf{y}, \hat{\mathbf{y}})=\min _{\mathbf{P}} \sum_{k=1}^{K_{\max}} \mathcal{L}_{\mathrm{BCE}}\left(\mathbf{y_k},[\mathbf{P} \mathbf{\hat{y}}]_k\right),
\]
where $\mathbf{\hat{y}}$ are the predicted speaker activations and $\mathbf{P}$ is an $K_{\max} \times K_{\max}$ permutation matrix and $\mathcal{L_{\mathrm{BCE}}}$ is the standard binary cross entropy loss.

Using the speaker annotations, we construct two audio chunks $(\mathbf{X^1},\mathbf{y^1})$ and $(\mathbf{X^2},\mathbf{y^2})$ with non-overlapping sets of speakers with the total number of speakers no greater than $K_{\max}$. Limiting the total number of speakers is critical in solving the over-separation issue of MixIT. The MoM is constructed as $\mathbf{X^{\mathrm{MoM}}} = \mathbf{X^1} + \mathbf{X^2} $ and the corresponding speaker activity labels $\mathbf{y^{\textrm{MoM}}}$ are given by $y_{:,t}^{\textrm{MoM}} =(y_{:,t}^{1}, y_{:,t}^{2})$ where the rows corresponding to non-active speakers are removed so that $\mathbf{y}^{\textrm{MoM}} \in\{0,1\}^{K_{\max }\times T}$. Then the MixIT loss function is given by,

\[\mathcal{L}_{\textrm{MixIT}}\left(\left\{\mathbf{X_n}\right\}, \hat{\mathbf{s}}\right)=\min _{\mathbf{A}} \sum_{n=1}^2 \mathcal{L_{\textrm{SI-SDR}}}\left(\mathbf{X_n},[\mathbf{A} \mathbf{\hat{s}}]_n\right),\]

\noindent where $\hat{\mathbf{s}}$ are the predicted separated sources, $M$ is the number of output sources and $\mathbf{A}$ is a mixing matrix $\mathbf{A} \in \{0,1\}^{2 \times M}$ under the constraint that each column sums to 1 and $\mathcal{L_{\textrm{SI-SDR}}}$ is the negative scale-invariant signal-to-distortion ratio \cite{rouxSDRHalfbakedWell2018}. Thanks to how we limit the total number of speakers when sampling the mixtures, we are able to use a significantly lower value for $M$.

Our combined multi-task loss is,
\begin{equation*}
\begin{split}
\mathcal{L}_{\mathrm{PixIT}} = \lambda \Big(\mathcal{L}_{\mathrm{PIT}}(\mathbf{y^1}, {\mathbf{\hat{y}^1}}) + \mathcal{L}_{\mathrm{PIT}}(\mathbf{y^2}, \mathbf{\hat{y}^2}) \\ + \mathcal{L}_{\mathrm{PIT}}(\mathbf{y^{\textrm{MoM}}}, {\mathbf{\hat{y}^{\textrm{MoM}}}})\Big) + (1-\lambda) \mathcal{L}_{\mathrm{MixIT}}\left(\left\{\mathbf{X_n}\right\}, \hat{\mathbf{s}}\right),
\end{split}
\end{equation*}

\noindent where among the three values, 0.1, 0.5, and 0.9, $\lambda=0.5$ was selected due to its superior performance on the development data. 

\subsection{Inference}

\begin{figure}[ht!]
  \centering
  \includegraphics[width=0.99\linewidth]{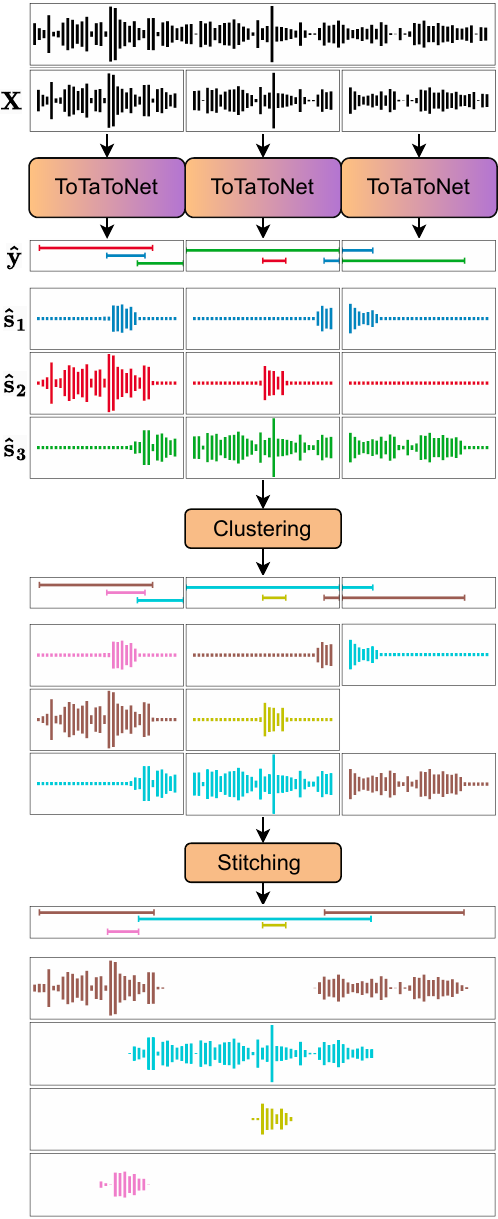}
  \caption{Inference on long-form audio. For ease of visualization inference using non-overlapping sliding windows is shown.}
  \label{fig:inference}
\end{figure}

During inference, an audio stream is partitioned into shorter chunks as depicted in Figure \ref{fig:inference}. The joint model processes each chunk and outputs aligned estimates for speaker sources and speaker activations. The resulting speaker activations and corresponding sources are clustered as in \cite{bredinPyannoteAudioSpeaker2023}. First, speaker activations are binarized using a detection threshold $\theta \in [0,1]$ to obtain speaker segments. Second, local speaker embeddings are extracted from each chunk for all the active speakers. We only utilize the regions of the chunk where the corresponding speaker is active. Speaker embeddings are computed by feeding the concatenation of original audio samples corresponding to those regions to the pre-trained ECAPA-TDNN model \cite{desplanques2020ecapa} available in \cite{ravanelli2021speechbrain}. Finally, agglomerative hierarchical clustering is performed on these embeddings using a clustering threshold $\delta$. As an important post-processing step, we perform leakage removal by setting the stitched separated sources at time $t$ to zero when the diarization outputs predict that the corresponding speaker is not active and has not been active in a window $[t-\Delta t,t+\Delta t]$. This is a key benefit of the aligned speaker activations and speaker sources since it eliminates all cross-talk when the corresponding speaker is not active.  The goal of introducing $\Delta t$ is to give downstream ASR systems additional context. The hyperparameters $\theta$, $\delta$, and $\Delta t$ are optimized on the development dataset.

\section{Experiments}

\subsection{Datasets}
\label{sec:data}
We chose two publicly-available real-world meeting datasets AMI and AliMeeting for our experiments. AMI \cite{carletta2005ami} consists of roughly 100 hours of English data.  AliMeeting \cite{yu_m2met_2022} is a Mandarin Chinese dataset with approximately 120 hours of recordings. As our goal is single-channel speech separation we only use the first channel of the microphone array also known as the single distant microphone (SDM) audio from AMI and channel 1 from AliMeeting for our experiments. Table \ref{tab:stats} shows statistics for the two datasets \cite{raj_gpu-accelerated_2023}. While both datasets consist of meeting recordings, AliMeeting contains significantly more overlap. In all our experiments, the ToTaToNet model is trained only on the train set of the corresponding dataset.

\begin{table}[b]
\centering
\caption{Statistics of datasets used for evaluations. The $k$-speaker durations are in terms of fraction of total speaking time.}
\label{tab:stats}
\adjustbox{max width=\linewidth}{
\begin{tabular}{@{}lcccccc@{}}
\toprule
 & \multicolumn{3}{c}{AMI} & \multicolumn{3}{c}{AliMeeting} \\
\cmidrule(lr{2pt}){2-4} \cmidrule(l{2pt}){5-7}
 & Train & Dev & Test & Train & Eval & Test \\ 
\midrule
Duration (h:m) & 79:23 & 9:40 & 9:03 & 111:21 & 4:12 & 10:46 \\
Num. sessions & 133 & 18 & 16 & 209 & 8 & 20 \\
Silence (\%) & 18.1 & 21.5 & 19.6 & 7.11 & 7.7 & 8.0 \\
1-speaker (\%) & 75.5 & 74.3 & 73.0 & 52.5 & 62.1 & 63.4 \\
2-speaker (\%) & 21.1 & 22.2 & 21.0 & 32.8 & 27.6 & 24.9 \\
$>$2-speaker (\%) & 3.4 & 3.5 & 6.0 & 14.7 & 10.2 & 11.7 \\
\bottomrule
\end{tabular}}
\end{table}

\subsection{Evaluation}

\textbf{Metrics.} As ground-truth reference sources are not available for real-world data we use ASR performance as a proxy for evaluating the quality of long-form separation. We apply an ASR system independently on each of the separated sources and report the word error rate (WER) between the speaker-attributed predictions and references. Multiple definitions of WER have been proposed for ASR systems that process audio with multiple speakers and output multiple word sequences (MIMO) \cite{vonneumannWordErrorRate2023}. We choose concatenated minimum permutation WER (cpWER) \cite{watanabe2020chime} as our main metric because it is the only one that penalizes speaker confusion which is unwanted for long-form speaker sources. On Mandarin data, this metric corresponds to the concatenated minimum-permutation character error rate (cpCER). We use the same text normalizer as Whisper for both English and Mandarin.

It is important to note that the definition of cpWER that we used does not penalize redundant hypothesis speaker channels \cite{vonneumannWordErrorRate2023}. For transparency we include results using a definition of cpWER that includes this penalization using MeetEval's implementation \cite{von2023meeteval}. These can be found in Appendix \ref{app:cpWER}.

On English data, we also report the utterance-wise WER (uWER) which ignores speaker attribution. The uWERs are calculated using Kaldi scripts \cite{poveyKaldiSpeechRecognition} which in turn utilize asclite \cite{fiscus2006multiple}. When using the long-form separated sources as input the single transcript is generated by first concatenating the ASR predictions from all the long-form sources and then sorting the words by start time.

 Our metric for evaluating the diarization performance is the diarization error rate (DER)~\cite{fiscus2006rich}, which is defined as the sum of false alarm, missed detection, and speaker confusion rates. No forgiveness collar is used.\\

\noindent\textbf{ASR systems.} To verify the quality of the separated sources we experiment with multiple ASR systems. For English data, we chose the small.en, medium.en, and large-v2 Whisper models \cite{radford_robust_2022} and NVIDIA's stt\_en\_conformer\_ctc\_large available in the NeMo toolkit \cite{kuchaiev2019nemo} on the basis that they were among the top performers on AMI as indicated by \cite{open-asr-leaderboard}. On Mandarin data, we only tested the aforementioned Whisper models with the English-only variants replaced with multilingual ones.\\

\noindent\textbf{Speaker attribution.} When evaluating cpWER (or cpCER for Mandarin AliMeeting), we compare two methods of adding speaker attribution (SA) to an ASR system. One through long-form separated sources and the other through speaker diarization. In the first case, the ASR systems are applied on the long-form separated sources immediately yielding speaker-attributed transcripts. In the latter, an ASR system is used on the original audio, and the predicted utterances are divided between speakers according to a speaker diarization system. Namely, each utterance is attributed to the speaker whose speaking segments have the most overlap with it. In the rare case that multiple speakers have fully overlapping speaking segments with the utterance, it is randomly attributed to one of them. In the following, we will refer to these two approaches as SA methods and refer to the system that was used to perform either diarization or separation as the SA system.\\

\noindent\textbf{Word timestamps.} For experiments with the Whisper family of models, we utilized WhisperX \cite{bain2023whisperx} which has implemented word-level time-stamps using forced alignment with a wav2vec2.0-based phoneme model \cite{baevski2020wav2vec}. The NeMo toolkit also provides word-level timestamps for the stt\_en\_conformer\_ctc\_large model. \\

\noindent{\textbf{Baselines}.} Our baseline systems perform speaker attribution through the \textit{pyannote.audio} 3.1 speaker diarization pipeline \cite{plaquetPowersetMulticlassCross2023}. 

\subsection{Implementation details}

During training, we sample the first mixture randomly across all the annotated regions from all the training files. Then we sample the second mixture from the same file while ensuring that it has no speakers in common with the first mixture and the total number of speakers is not greater than the number of output sources of the model. Sampling the other chunk from the same file has two benefits. First, it is important that the two mixtures come from the same recording conditions, otherwise the model might learn to exploit this difference as found in \cite{sivaramanAdaptingSpeechSeparation2021}. Second, this approach generalizes better because it does not require dataset-wise consistent speaker IDs.

Our system is implemented in the \textit{pyannote.audio} toolkit \cite{bredinPyannoteAudioSpeaker2023} with the help of the \textit{Asteroid} library~\cite{pariente_asteroid_2020}. We use 5-second sliding windows with a step size of 500ms as in \cite{bredinPyannoteAudioSpeaker2023} and in line with \cite{sivaramanAdaptingSpeechSeparation2021}. For both AMI and AliMeeting, there is a less than 1\% chance that a 5-second window contains more than three active speakers~\cite{plaquetPowersetMulticlassCross2023}. Motivated by this statistic and aiming to mitigate over-separation, we set $K_{\max}=3$. As a consequence of our sampling method for the mixtures, training data does not include windows with more than three speakers.

In ToTaToNet, the 1D conv encoder and decoder use a kernel size of 32, a stride of 16, and 64 filters. We concatenate the encoder output with WavLM-large pre-trained features which have a stride of 320 so the WavLM features are repeated 20 times. For the separator module we chose a DPRNN \cite{luoDualpathRNNEfficient2020} with chunk size 100, hop size 50, and the rest of the hyperparameters kept the same as in the original work. The diarization module starts with an 8-fold average pooling layer to decrease the temporal resolution to that of \cite{bredinPyannoteAudioSpeaker2023}. We follow it with a simple diarization model consisting of a fully connected neural network with two 64-dimensional layers. We thus rely on the masking network to do the bulk of the work for speaker diarization. Importantly, due to the PIT training for diarization, the diarization module has to process each encoded masked source separately (as does the 1D conv decoder) otherwise the diarization outputs might be permuted with respect to the separated sources.

We use a learning rate of $1e^{-5}$ for the WavLM parameters and $3e^{-4}$ for the rest of the parameters. The learning rate is halved whenever the validation loss plateaus for 5 epochs. We use the Adam optimizer \cite{kingma2014adam} with the gradients clipped to a $L_2$-norm of 5 and train all models for 100 epochs. 

When optimizing for the hyperparameters $\Delta t$, $\theta$, and $\delta$, we used either cpWER/cpCER or DER as the target metric depending on whether the pipeline was used for separation or diarization.

To ensure reproducibility, the code for both training and inference using PixIT will be available in the open-source \textit{pyannote.audio} library. The recipes and separated source samples will be publicly available at {\scriptsize\url{github.com/joonaskalda/PixIT}}.

\subsection{Results}

The cpWERs for the various ASR systems on AMI-SDM test set are shown in Table \ref{tab:ami_cpwer}.  We can see that long-form separation via PixIT significantly improves the quality of speaker-attributed transcripts across the variety of ASR systems used. Notably, the ASR systems are applied on the separated sources off-the-shelf with no fine-tuning required.

\begin{table}[t]
\centering
\caption{The cpWER (\%) on AMI-SDM for various ASR systems with speaker attribution (SA) done through diarization or the joint model}\label{tab:ami_cpwer}
\setlength{\tabcolsep}{3pt}   
\adjustbox{max width=\linewidth}{
\begin{tabular}{lccccccl}
\toprule
\multirow{2}{*}{ASR model} & \multirow{2}{*}{SA method} & \multirow{2}{*}{SA system} & \multicolumn{4}{c}{cpWER(\%)} & \multirow{2}{*}{\makecell{Relative \\ Change}} \\
\cmidrule{4-7}
& & & sub & del & ins & \textbf{total} & \\
\midrule
\multirow{3}{*}{Whisper small.en} 
& Diarization & pyannote 3.1 & 8.7 & 27.2 & 3.7 & \textbf{39.6} & \\
& Diarization & PixIT & 8.5 & 27.3 & 2.1 & \textbf{37.9} & -4.3\% \\
& Separation & PixIT & 6.7 & 25.8 & 1.4 & \textbf{33.9} & -14.4\% \\
\cmidrule{1-8}
\multirow{3}{*}{Whisper medium.en} 
& Diarization & pyannote 3.1 & 7.4 & 28.0 & 3.4 & \textbf{38.8} & \\
& Diarization & PixIT & 7.3 & 27.8 & 2.0 & \textbf{37.1} & -4.4\% \\
& Separation & PixIT & 5.9 & 25.8 & 1.2 & \textbf{32.8} & -15.4\% \\
\cmidrule{1-8}
\multirow{3}{*}{Whisper large-v2} 
& Diarization & pyannote 3.1 & 7.1 & 29.3 & 1.8 & \textbf{38.3} & \\
& Diarization & PixIT & 6.9 & 26.6 & 2.1 & \textbf{35.7} & -6.7\% \\
& Separation & PixIT & 5.6 & 24.7 & 1.3 & \textbf{31.7} & -17.2\% \\
\cmidrule{1-8}
\multirow{3}{*}{NeMo conformer large} 
& Diarization & pyannote 3.1 & 11.5 & 36.0 & 1.4 & \textbf{48.9} & \\
& Diarization & PixIT & 13.3 & 33.9 & 1.3 & \textbf{48.5} & -0.8\% \\
& Separation & PixIT & 13.4 & 24.6 & 1.4 & \textbf{39.4} & -19.4\% \\
\bottomrule
\end{tabular}}
\end{table}

We also report the uWER scores using either the original audio or the separated sources in Table \ref{tab:ami_wer}. Across the ASR models, the bulk of the WER improvement comes from deletions. Having the original audio as input the ASR models may miss the quieter speakers utterances during overlap and utilizing separated sources helps recover those.

\begin{table}[t]\centering
\caption{The uWER (\%) on AMI-SDM for various ASR systems using either the original audio or the separated sources as input}\label{tab: ami_wer}
\setlength{\tabcolsep}{3pt}   
\adjustbox{max width=\linewidth}{
\begin{tabular}{lcccccc}\toprule
\multirow{2}{*}{ ASR model } & \multirow{2}{*}{Input to ASR} & \multicolumn{4}{c}{uWER(\%)}  & \multirow{2}{*}{\makecell{Relative \\ change}} \\\cmidrule{3-6}
& & sub & del & ins & \textbf{total} & \\\cmidrule{1-7}
\multirow{2}{*}{ Whisper small.en } & Original audio & 6.7 & 29.6 & 1.4 & \textbf{37.6} & \\
& Separated sources & 6.9 & 27.9 & 1.5 & \textbf{36.3} & -3.5\% \\\midrule
\multirow{2}{*}{ Whisper medium.en } & Original audio & 5.8 & 30.0 & 1.3 & \textbf{37.1} & \\
& Separated sources & 6.0 & 27.7 & 1.4 & \textbf{35.1} & -5.4\% \\\midrule
\multirow{2}{*}{ Whisper large-v2 } & Original audio & 5.2 & 28.9 & 1.3 & \textbf{35.4} & \\
& Separated sources & 5.5 & 26.9 & 1.4 & \textbf{33.8} & -4.5\% \\\midrule
\multirow{2}{*}{ NeMo conformer large } & Original audio & 10.7 & 36.7 & 1.8 & \textbf{49.3} & \\
& Separated sources & 12.6 & 26.4 & 2.6 & \textbf{41.6} & -15.6\% \\
\bottomrule
\end{tabular}}
\label{tab:ami_wer}
\end{table}

Table \ref{tab:alimeeting_cpwer} shows the cpCERs for the AliMeeting channel 1 dataset. We can see that the improvement from utilizing separated sources is greater than 20\% across the tested ASR systems. Notably, the relative improvements are greater than they were for the corresponding models on AMI data even though WavLM has been pre-trained on English data. This can be explained by the greater percentage of overlap present in AliMeeting as mentioned in section \ref{sec:data}.  

\begin{table}[t]\centering
\caption{The cpCER (\%) on Alimeeting channel 1 for various ASR systems with speaker attribution (SA) done through diarization or the joint model}
\setlength{\tabcolsep}{3pt}
\adjustbox{max width=\linewidth}{
\begin{tabular}{lccccccc}
\toprule
\multirow{2}{*}{ASR system} & \multirow{2}{*}{SA} & \multirow{2}{*}{SA model} & \multicolumn{4}{c}{cpCER(\%)} & \multirow{2}{*}{\makecell{Relative \\ Change}} \\
\cmidrule{4-7}
& & & sub & del & ins & \textbf{total} & \\
\midrule
\multirow{3}{*}{Whisper small} 
& Diarization & pyannote 3.1 & 23.4 & 35.6 & 9.6 & \textbf{68.6} & \\
& Diarization & PixIT & 23.3 & 35.1 & 9.5 & \textbf{67.9} & -1.0\% \\
& Separation & PixIT & 16.2 & 33.4 & 4.4 & \textbf{54.0} & -21.3\% \\
\midrule
\multirow{3}{*}{Whisper medium} 
& Diarization & pyannote 3.1 & 18.5 & 37.9 & 9.5 & \textbf{65.9} & \\
& Diarization & PixIT & 18.8 & 37.2 & 8.9 & \textbf{64.9} & -1.5\% \\
& Separation & PixIT & 11.8 & 34.2 & 4.2 & \textbf{50.3} & -23.7\% \\
\midrule
\multirow{3}{*}{Whisper large-v2} 
& Diarization & pyannote 3.1 & 17.6 & 38.0 & 9.5 & \textbf{65.1} & \\
& Diarization & PixIT & 18.1 & 37.3 & 9.0 & \textbf{64.4} & -1.1\% \\
& Separation & PixIT & 10.6 & 33.6 & 4.0 & \textbf{48.3} & -25.8\% \\
\bottomrule
\end{tabular}}
\label{tab:alimeeting_cpwer}
\end{table}

In Table \ref{tab:ablation}, we show the effects of adding the WavLM features and performing leakage removal through the diarization output on our system performance when performing SA-ASR on AMI-SDM with Whisper medium.en. The system without WavLM features clearly has issues with leakage, with a lot of it passing through the VAD component of WhisperX. When using WavLM features the effect of our leakage removal is smaller but it still outperforms using only WhisperX. Notably, a decrease in substitution errors from applying leakage removal can be observed in both cases. A possible explanation is that since the leakage removal reduces the length of the predicted text, some words in the reference that previously corresponded to substitution errors now count as deletion errors. This is verified by the fact that the decrease in substitution errors is smaller than the increase in deletion errors. This effective method for leakage removal is a further benefit of ToTaToNet's aligned outputs. Leveraging the WavLM features significantly improves our system's performance which makes sense given the relatively small amount of data we have access to for training and the utterance mixing component of the pre-training. Still, even without using pre-trained features, we are able to improve on the baseline of WhisperX.

\begin{table}[t]\centering
\caption{The cpWER (\%) on AMI-SDM for speaker-attribution (SA) done through PixIT speech separation with different configurations of WavLM and leakage removal. Whisper medium.en is used for ASR and pyannote 3.1 diarization as the baseline SA method.}\label{tab:ablation}
\setlength{\tabcolsep}{3pt}   
\adjustbox{max width=\linewidth}{
\begin{tabular}{lccccccc}\toprule
 \multirow{2}{*}{SA method} & \multirow{2}{*}{WavLM} & \multirow{2}{*}{\makecell{Leakage\\ removal}} & \multicolumn{4}{c}{cpWER(\%)} & \multirow{2}{*}{\makecell{Relative \\ change}} \\\cmidrule{4-7}
& & & sub & del & ins & \textbf{total} & \\\cmidrule{1-8}
pyannote 3.1 & & & 7.4 & 28.0 & 3.4 & \textbf{38.8} & \\\cmidrule{1-8}
\multirow{4}{*}{PixIT separation} & \xmark & \xmark & 19.2 & 15.3 & 15.6 & \textbf{50.1} & +29.1\%\\
& \xmark & \cmark & 6.4 & 28.1 & 1.7 & \textbf{36.2} & -6.7\% \\
& \cmark & \xmark & 9.3 & 21.0 & 3.8 & \textbf{34.1} & -12.1\% \\
& \cmark & \cmark & 5.9 & 25.8 & 1.2 & \textbf{32.8} & -15.5\% \\
\bottomrule
\end{tabular}}
\end{table}

We also analyze PixIT's speaker diarization performance by measuring the DERs on AMI-SDM and AliMeeting for various training and hyperparameter optimization strategies as shown in Table \ref{tab:der}. For the systems optimized for cpWER, we use $\Delta t = 0$, as that represents the real diarization capabilities. We have included the state-of-the-art (SOTA) systems as of February 2024. For AliMeeting this is the pyannote 3.1 system utilizing powerset training \cite{plaquetPowersetMulticlassCross2023} and for AMI-SDM it is the end-to-end diarization model leveraging the Mask2Former architecture proposed in \cite{harkonen2024eend}. The DER scores are broken down into false alarm (FA), missed detection (MD), and speaker confusion (SC) rates. Optimizing for cpWER yields lower FA values. This means that a higher speaker activation threshold $\theta$ is used and only segments for which the diarization branch is confident are considered for ASR. Optimizing the $\lambda = 0.5$ system for DER improves on the SOTA for both AliMeeting and AMI-SDM. Training our system for only the easier task of speaker diarization i.e. with $\lambda = 1$, we achieve a further boost to performance on both datasets.

\begin{table}[t]
\centering
\caption{DER (\%) comparison with state-of-the-art systems on AMI-SDM and AliMeeting channel 1 for different training strategies and ways of optimizing the hyperparameters $\theta$, $\delta$, and $\Delta t$. For the latter, the underlying ToTaToNet is kept the same.}\label{tab:der}
\setlength{\tabcolsep}{3pt}   
\adjustbox{max width=\linewidth}{
\begin{tabular}{lcccc}
\toprule
& \multicolumn{4}{c}{DER(\%)} \\\cmidrule{2-5}
AMI-SDM systems & FA & MD & SC & \textbf{total} \\
\midrule
H{\"a}rk{\"o}nen et al. \cite{harkonen2024eend} &  &  &  & \textbf{18.9} \\\midrule
PixIT, $\lambda=0.5$, optimized for cpWER & 1.3 & 17.9 & 6 & \textbf{25.3} \\
PixIT, $\lambda=0.5$, optimized for DER & 3.9 & 8.2 & 5.6 & \textbf{17.7} \\
PixIT, $\lambda=1$ & 4.4 & 7.2 & 5.5 & \textbf{17.1} \\\toprule
AliMeeting systems & &  & & \\\midrule
Plaquet et al. \cite{plaquetPowersetMulticlassCross2023} & 3.7 & 10.4 & 9.2 & \textbf{23.3} \\\midrule
PixIT, $\lambda=0.5$, optimized for cpWER & 2.7 & 13.2 & 12.4 & \textbf{28.3} \\
PixIT, $\lambda=0.5$, optimized for DER & 5.8 & 7.3 & 8.3 & \textbf{21.4} \\
PixIT, $\lambda=1$ & 4.7 & 6.5 & 8.3 & \textbf{19.5} \\
\bottomrule
\end{tabular}}
\end{table}

\section{Conclusion}

In this paper, we proposed PixIT, a novel approach for performing multitask training for speaker diarization and speech separation. This method does not depend on clean single-speaker individual sources, only requiring single-channel recordings with speaker diarization labels which are usually a part of annotation. The local separated source and diarization predictions of the proposed ToTaToNet model are aligned allowing for long-form inference via the \textit{best-of-both-worlds} approaches that have been developed for speaker diarization. A further benefit of the aligned sources is that we can perform effective leakage removal by zeroing out inactive speaker sources. We perform various experiments to demonstrate the quality of the long-form separated sources obtained from real-world meeting data by using them as input for various ASR systems. Indeed, the cpWERs show significant improvements over the baseline of performing speaker attribution using speaker diarization with the improvements increasing with the proportion of overlapped speech present. Furthermore, we observe a decrease in utterance-based WER when the ASR outputs from separated sources are combined into a single transcript. These results come from using the ASR systems on the separated sources off the shelf with no fine-tuning required. Finally, we show that PixIT achieves state-of-the-art speaker diarization performance on both the AMI-SDM and AliMeeting datasets.

\section{Acknowledgements}

We would like to acknowledge Dr Yoshiaki Bando, for pointing out the ambiguity in cpWER definitions regarding scenarios involving an unknown number of speakers. The research reported in this paper was supported by the Agence de l'Innovation Défense under the grant number 2022~65~0079.
This work was granted access to the HPC resources of GENCI-IDRIS under the allocations AD011014274, as well as the TalTech supercomputing resources.

\bibliographystyle{IEEEbib}
\bibliography{refs}

\clearpage

\appendix 

\section{On the evolution of cpWER definitions}
\label{app:cpWER}

\subsection{Review of cpWER definitions}

Here we provide a brief literature review on cpWER definitions as of early 2023, which served as the basis for our choice of a cpWER variant that does not penalize overestimation.

The original cpWER definition, as proposed in CHiME-6, was limited to scenarios involving a fixed number of speakers \cite{watanabe2020chime}. In our work, we adopted the extended cpWER definition from \cite{vonneumannWordErrorRate2023}, which offered a comprehensive review of multi-speaker word error rate definitions. Given the clarity and extensiveness of their approach, it seemed appropriate for us to follow their methodology. In this definition, underestimation of speakers is penalized by adding empty dummy channels to the hypothesis, whereas overestimation is not penalized. However, in a subsequent paper \cite{von2023meeteval}, the authors revised their definition to also penalize overestimation by adding dummy channels to the reference.

In other papers dealing with scenarios involving an unknown number of speakers, the definitions were often ambiguous, or they indicated that redundant hypothesis speakers were discarded. For instance, \cite{liSelfSupervisedLearningBasedSource2023} and \cite{zheng2022tandem} mention the removal of redundant speakers, denoting this variant as cpWER-us when dealing with an unknown number of speakers.

In the series of papers on Serialized Output Training by Naoyuki Kanda et al., the initial paper, \cite{kandaSerializedOutputTraining2020}, touches on the problem of having more hypothesis speakers than references but is vague about the exact resolution (referring to the metric as WER, though effectively it aligns with cpWER). Later works in this series did not provide further clarification on this issue.

\subsection{Results using the MeetEval cpWER definition}

For completeness, we provide the results calculated using the cpWER definition that penalizes overestimation, utilizing the MeetEval toolkit \cite{von2023meeteval}. The inference hyperparameters $\theta$, $\delta$, and $\Delta t$ were re-optimized on the development dataset using this cpWER definition. Results based on this updated cpWER can be found in Tables \ref{tab:ami_cpwer_alt} and \ref{tab:alimeeting_cpwer_alt} for AMI-SDM and Alimeeting channel 1, respectively. The relative changes in cpWER are consistent with those obtained using the original definition.

\begin{table}[H]
  \centering
  \caption{MeetEval cpWER (\%) results on AMI-SDM for various ASR models with speaker attribution (SA) through diarization or separation.}
  \setlength{\tabcolsep}{3pt}   
  \adjustbox{max width=\linewidth}{
  \begin{tabular}{lccccccc}
  \toprule
  \multirow{2}{*}{ASR model} & \multirow{2}{*}{SA method} & \multirow{2}{*}{SA system} & \multicolumn{4}{c}{cpWER(\%)} & \multirow{2}{*}{\makecell{Relative \\ Change}} \\
  \cmidrule{4-7}
  & & & sub & del & ins & \textbf{total} & \\
  \midrule
  \multirow{3}{*}{Whisper small.en} 
  & Diarization & pyannote 3.1 & 7.6 & 29.0 & 4.0 & \textbf{40.5} & \\
  & Diarization & ToTaToNet & 7.8 & 27.2 & 2.2 & \textbf{37.2} & -8.1\% \\
  & Separation & ToTaToNet & 8.8 & 24.2 & 2.4 & \textbf{35.4} & -12.6\% \\
  \cmidrule{1-8}
  \multirow{3}{*}{Whisper medium.en} 
  & Diarization & pyannote 3.1 & 6.7 & 29.7 & 3.6 & \textbf{40.0} & \\
  & Diarization & ToTaToNet & 7.0 & 28.1 & 2.0 & \textbf{37.1} & -7.3\% \\
  & Separation & ToTaToNet & 7.6 & 24.1 & 2.2 & \textbf{33.9} & -15.3\% \\
  \cmidrule{1-8}
  \multirow{3}{*}{Whisper large-v2} 
  & Diarization & pyannote 3.1 & 6.4 & 28.0 & 3.9 & \textbf{38.3} & \\
  & Diarization & ToTaToNet & 6.8 & 26.3 & 2.1 & \textbf{35.2} & -8.1\% \\
  & Separation & ToTaToNet & 7.3 & 22.7 & 2.6 & \textbf{32.6} & -14.9\% \\
  \cmidrule{1-8}
  \multirow{3}{*}{Nemo conformer large} 
  & Diarization & pyannote 3.1 & 12.0 & 35.5 & 2.9 & \textbf{50.4} & \\
  & Diarization & ToTaToNet & 13.2 & 34.1 & 1.6 & \textbf{48.9} & -3.0\% \\
  & Separation & ToTaToNet & 15.7 & 23.7 & 2.0 & \textbf{41.4} & -17.9\% \\
  \bottomrule
  \end{tabular}}
  \label{tab:ami_cpwer_alt}
  \end{table}

  \begin{table}[H]
    \centering
    \caption{MeetEval cpCER (\%) results on Alimeeting channel 1 for various ASR models with speaker attribution (SA) through diarization or separation.}
    \setlength{\tabcolsep}{3pt}   
    \adjustbox{max width=\linewidth}{
    \begin{tabular}{lccccccc}
    \toprule
    \multirow{2}{*}{ASR model} & \multirow{2}{*}{SA method} & \multirow{2}{*}{SA system} & \multicolumn{4}{c}{cpCER(\%)} & \multirow{2}{*}{\makecell{Relative \\ Change}} \\
    \cmidrule{4-7}
    & & & sub & del & ins & \textbf{total} & \\
    \midrule
    \multirow{3}{*}{Whisper small} 
    & Diarization & pyannote 3.1 & 23.2 & 35.4 & 10.0 & \textbf{68.6} & \\
    & Diarization & ToTaToNet & 23.1 & 35.0 & 9.6 & \textbf{67.7} & -1.3\% \\
    & Separation & ToTaToNet & 18.9 & 32.4 & 2.4 & \textbf{53.7} & -21.7\% \\
    \cmidrule{1-8}
    \multirow{3}{*}{Whisper medium} 
    & Diarization & pyannote 3.1 & 18.5 & 37.9 & 9.5 & \textbf{65.9} & \\
    & Diarization & ToTaToNet & 18.7 & 37.3 & 8.9 & \textbf{64.9} & -1.5\% \\
    & Separation & ToTaToNet & 12.5 & 34.7 & 1.6 & \textbf{48.7} & -26.1\% \\
    \cmidrule{1-8}
    \multirow{3}{*}{Whisper large-v2} 
    & Diarization & pyannote 3.1 & 17.9 & 37.9 & 9.9 & \textbf{65.6} & \\
    & Diarization & ToTaToNet & 18.1 & 37.3 & 9.3 & \textbf{64.7} & -1.4\% \\
    & Separation & ToTaToNet & 13.0 & 32.5 & 1.8 & \textbf{47.3} & -27.9\% \\
    \bottomrule
    \end{tabular}}
    \label{tab:alimeeting_cpwer_alt}
    \end{table}

\end{document}